PERSPECTIVE

# Intelligent Soft Matter: Towards Embodied Intelligence

Vladimir A. Baulin[a,b,*], Achille Giacometti[c], Dmitry A. Fedosov[d], Stephen Ebbens[e], Nydia R. Varela-Rosales[f], Neus Feliu[g], Mithun Chowdhury[h], Minghan Hu[i], Rudolf Füchslin[j], Marjolein Dijkstra[k], Matan Mussel[l], René van Roij[m], Dong Xie[n], Vassil Tzanov[o], Mengjie Zu[p], Samuel Hidalgo-Caballero[q], Ye Yuan[r], Luca Cocconi[s], Cheol-Min Ghim[t], Cécile Cottin-Bizonne[u], M. Carmen Miguel[v], Maria Jose Esplandiu[w], Juliane Simmchen[x], Wolfgang J. Parak[y], Marco Werner[z], Gerhard Gompper[aa], Martin M. Hanczyc[bb,cc]

Intelligent soft matter lies at the intersection of materials science, physics, and cognitive science, promising to change how we design and interact with materials. This transformative field aims to create materials with life-like capabilities, such as perception, learning, memory, and adaptive behavior. Unlike traditional materials, which typically perform static or predefined functions, intelligent soft matter can dynamically interact with its environment, integrating multiple sensory inputs, retaining past experiences, and making decisions to optimize its responses. Inspired by biological systems, these materials leverage the inherent properties of soft matter such as flexibility, adaptability, and responsiveness to perform functions that mimic cognitive processes. By synthesizing current research trends and projecting their evolution, we present a forward-looking perspective on how intelligent soft matter could be constructed, with the aim of inspiring innovations in areas such as biomedical devices, adaptive robotics, and beyond. We highlight new pathways for integrating sensing, memory and actuation with low-power internal operations, and we discuss key challenges in realizing materials that exhibit truly "intelligent behavior". These approaches outline a path toward more robust, versatile, and scalable materials that can potentially act, compute, and "think" through their inherent intrinsic material properties--moving beyond traditional smart technologies that rely on external control.

a. Universitat Rovira i Virgili, Tarragona, Spain
b. Active Inference Institute, Davis, California, USA
c. Dipartimento di Scienze Molecolari e Nanosistemi, Universita` Ca' Foscari Venezia, Via Torino 155, 30172 Venezia, Italy European Centre for Living Technology (ECLT) Ca' Bottacin, Dorsoduro 3911, Calle Crosera, 30123 Venice, Italy.
d. Forschungszentrum Juelich Germany
e. Department of Chemical and Biological Engineering University of Sheffield Sheffield S1 3JD, UK
f. AIST-AIMR, Tohoku University, Sendai 980-8577, Japan.
g. Zentrum für Angewandte Nanotechnologie CAN, Fraunhofer- Institut für Angewandte Polymerforschung IAP, Hamburg, Germany.
h. Indian Institute of Technology Bombay, Mumbai-400 076, India
i. Institute of Robotics and Intelligent Systems (IRIS), Swiss Federal Institute of Technology (ETH) Zurich, Switzerland
j. Zurich University of Applied Sciences, Zurich, Switzerland
k. Debye Institute for Nanomaterials Science, Utrecht University, Utrecht, the Netherlands
l. Department of Physics, Faculty of Natural Sciences, University of Haifa, Haifa, Israel
m. Institute for Theoretical Physics, Utrecht University, The Netherlands
n. University of Edinburgh, Edinburgh, UK
o. Universitat Rovira i Virgili, Tarragona, Spain
p. Institute of Science and Technology, Vienna, Austria
q. ESPCI, Paris, France
r. Hiroshima University, Japan
s. Max Planck Institute for Dynamics and Self-Organization, Göttingen, Germany
t. Departments of Physics and Biomedical Engineering, Ulsan National Institute of Science and Technology, Ulsan 44919, South Korea
u. University Lyon, Lyon, France
v. Universitat de Barcelona, Barcelona, Spain
w. Catalan Institute of Nanoscience and Nanotechnology (ICN2), CSIC and BIST, Campus UAB, Bellaterra,Barcelona
x. University of Strathclyde, Glasgow, UK
y. Universität Hamburg, Hamburg, Germany
z. Leibniz-Institut für Polymerforschung Dresden e.V., Dresden, Germany
aa. Forschungszentrum Juelich Germany
bb. Department of Cellular, Computational and Integrative Biology (CIBIO), University of Trento, Via Sommarive 9, Povo, Trento 38123, Italy.
cc. Department of Chemical and Biological Engineering, University of New Mexico, Albuquerque, NM 87106, USA.

## Introduction

The field of Intelligent Soft Matter (ISM) aims to develop materials with advanced capabilities, such as perception, memory, learning, and adaptive responses---features traditionally attributed only to living organisms. The transformative potential of ISM lies in its ability to autonomously interact with and adapt to its environment through self-learning and adaptive mechanisms that harness the inherent stochasticity of large numbers of interacting units[1]. ISM holds the promise of enabling a wide range of applications, ranging from sentient nanoscale robotics capable of navigating complex biological environments[2,3] to dynamically adaptive systems that optimize their performance in real-time[4,5], to materials that not only sense but also actively modify their surroundings, and to neuromorphic platforms for material-based cognition[6] and thermodynamic computing[7–9]. As such it requires combining aspects of materials science, physics, cognitive science, and engineering.

While the initial development of ISM is rooted in the quest to mimic the complexity and efficiency of biological systems such as neural networks[10–13], its potential may go well beyond mere biomimetics. The ultimate vision is to create soft matter agents with fully synthetic intelligence, capable of functionalities that may not even have biological analogues. Similar to macroscale soft robotics[1,14], intelligent soft matter should go significantly beyond passive adaptability towards more sophisticated levels of material intelligence, characterized by emergent behaviour and evolving functionalities in response to environmental cues[15]. Emergent agency in such systems[16] arises when the collective interactions of numerous nanoscale components give rise to system-level behaviours that exhibit goal-directed actions[17] or decision-making capabilities that are absent in the individual constituents. This sophisticated adaptivity is fundamentally rooted in the stochastic reorganization of micro- and nanoscale components[18,19], allowing for the probabilistic emergence and evolution of functional capabilities in response to environmental stimuli. For instance, self-assembling peptide structures exhibit dynamic morphological changes in response to localized chemical signals[20], demonstrating a sophisticated level of responsiveness that can be harnessed for molecular information processing.

Current research in intelligent matter showcases the potential for cognitive-like behaviours, primarily demonstrated through macroscopic systems, such as shape-changing robots and self-healing materials[7,14]. However, these macroscopic approaches often rely on a separation of functional layers: sensing, memory, and actuation[21], which poses severe limitations when scaled down to the microscale[22]. At smaller dimension, the intricate orchestration and communication required between numerous specialized components become increasingly inefficient due to the high number of interactions and the growing influence of interfacial effects[23]. Thus, the future of intelligent soft matter lies in constructing systems from a vast number of similar, interacting units where complex functionalities, such as sensing, memory, and actuation, emerge intrinsically and are distributed throughout the material. These systems should not require intricate hierarchical control or the integration of distinct, specialized components. Instead, their emergent behaviour will additionally capitalize on intrinsic stochasticity and its manifestations, such as thermodynamic fluctuations, phase transitions, and self-assembly processes.

**Specific examples:** Realizing this vision requires the design of materials where function is inherently integrated in a distributed way rather than externally added on. For example,

- The behaviour of active liquid crystals exhibiting spontaneous flows and pattern formation can be viewed through the lens of stochasticity leading to emergent computation[24]. Another work[25] illustrates how cholesteric liquid crystal fingers can perform geometric and logical computations. Utilizing voltage-controlled reorientation in thin films of liquid crystals, the presented system achieves computational tasks through the manipulation of topological defects. Specifically, the authors showcase the material's capacity for approximating Voronoi diagrams and implementing one-bit half-adder logic through engineered defect collisions. More broadly, the large-scale self-organization of topological defect networks in nematic liquid crystals can be manipulated to create stable, periodic patterns[26] showing potential for distributed sensing and information processing.
- Networks of responsive nanoparticles exhibit remarkable capabilities to sense changes in their surroundings and anticipate future states based on past experiences, enabling proactive responses. Thus, a nanoparticle-based computing architecture that utilizes nanoparticles as hardware and DNA strands as software can be used to create programmable logic circuits[27]. This architecture allows for the formation of nanoparticle neural networks capable of performing complex computations, including Boolean logic operations. Analogously, a nanograin network memory device[28] that utilizes reconfigurable percolation paths can exhibit synaptic behaviours, such as potentiation and habituation.
- Self-assembling polymers can adapt locally for distributed responses and memory[29], allowing a single detection event to trigger widespread material changes, such as switching from hydrophobic to hydrophilic states. Similarly, nucleic acids structures offer a promising avenue for programming distributed information processing.
- Dynamic colloidal assemblies, such as magnetic droplets at the air-liquid interface[30], form highly ordered patterns that can be precisely controlled using external magnetic fields. Similarly, acoustic signalling can similarly enable collective perception and control in active particles[31] showing how local actions can create overall control within a system. These assemblies exemplify adaptive materials capable of computation and responsiveness to their environment.
- A nanopore interface for higher-bandwidth DNA computing[32], can be used for real-time data processing and enhances the throughput of DNA-based computational systems.

Such distributed systems, composed of vast numbers of interacting units, promise ISM systems with unprecedented levels of autonomy, decentralization, and adaptability, blurring the traditional distinction between inorganic substrates and living systems. Thermodynamic machines with predictive capabilities draw inspiration from Maxwell's demon – a conceptual device that highlights how information can be utilized to seemingly circumvent thermodynamic limitations, rather than to violate them directly. The rigorous framework of "information thermodynamics" addresses Maxwell's demon paradox by explicitly quantifying the thermodynamic costs of measurement and information erasure[33], therefore preserving the second law of thermodynamics while also providing fundamental limits for device performance and operational energy expenditure[34–36].

Boltzmann machines are an interesting theoretical concept[37] that illustrate how entropic contributions in a magnetic system with fine-tuned couplings between spins could be exploited to generalize scattered input data – and generate new random and plausible data examples with a similar distribution. Here, the right balance between susceptibility for new input patterns and the minimal memory required for their generalization have a traceable route to the thermodynamics of phase transitions in magnetic systems[38,39]. Boltzmann machines are different from feed forward artificial neural networks through their bidirectionality - a feature that allows for recurrence and the internal, iterative restructuring of input information. Alone, experimental realizations of this mathematical model currently seem unrealistic.

While the physical intelligence paradigm represents a significant step towards harnessing material physics, many current examples, including those reviewed under the physical intelligence umbrella[6] often focus on physically implementing functions specified by human design. This includes realization of logic gates[40,41], storage of predefined states[42], execution of pre-programmed responses[43], or the use of external control systems to orchestrate complex behaviours[44,45].

This perspective seeks to build upon and refine the concept of intelligence in materials by focusing specifically on emergent material intelligence. We argue for a deeper integration in which intelligent behaviors emerge fundamentally and autonomously from the material's intrinsic physics. This approach leverages nonlinearity, stochasticity, self-organization, and adaptive feedback operating far from equilibrium. Rather than relying primarily on externally imposed designs or algorithms within a physical substrate, intelligence should arise naturally from these inherent properties. The emphasis shifts from physical encoding of designed function towards physical generation of emergent function.

While this vision of nanoscale material intelligence is compelling, its practical realization presents profound scientific and engineering challenges: how can we rationally design material architectures that effectively adapt to dynamically changing environments? How can we reliably harness the principles of stochasticity and the collective behaviour of vast numbers of nanoscale replicas to achieve complex behaviour with prediction? How do we imbue materials with the capacity to learn from past interactions and make autonomous decisions based on their accumulated experiences and real-time perception? This perspective addresses these critical challenges, outlining a roadmap for the future of ISM.

## Main Capabilities of Intelligent Soft Matter

For soft matter system to exhibit material intelligence, it must possess a minimal level of structural complexity (Fig. 1).

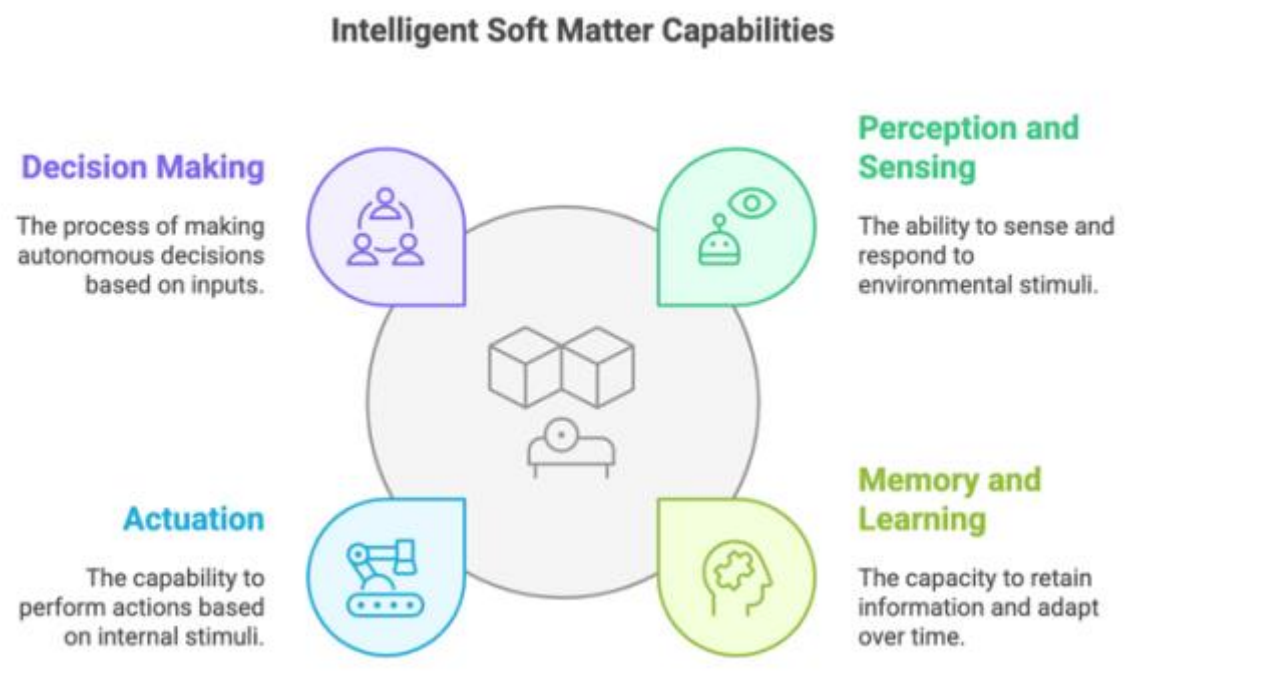

Figure 1 Four key capabilities essential for Intelligent Soft Matter.

### 1. Perception and Sensing

To respond adaptively, an ISM system must sense its environment, interpret stimuli, and prioritize actions accordingly.

Desirable functions and capabilities include:

**(i) Global Response Integration**: ISM should be capable of processing multiple types of sensory inputs, such as pressure, temperature, or chemical gradients, and synthesizing these into a coherent response. For example, soft robots that utilize continuous fluid flow for integrated control can process multiple sensory inputs and respond effectively, achieving robust control without complex discrete valving[46].

**(ii) Selective Attention/Filtering:** This feature enables the material to selectively respond to certain stimuli while filtering out irrelevant signals or noise, focusing on significant environmental changes. Neuromorphic principles realized in physical networks, like criticality, are hypothesized to optimize sensitivity and dynamic range, effectively filtering information based on relevance[47,48]. Physical systems exhibiting thresholding, like damage-indicating microcapsules[49,50], perform basic filtering based on stimulus intensity.

**(iii) Dynamic Sensory-Actuator Coupling:** ISM requires tight, dynamic coupling between sensing and actuation via internal feedback loops. This allows for continuous monitoring and iterative adjustment, modulating responses based on ongoing interactions. While many current systems separate sensing and actuation[51], materials exhibiting self-regulation, like homeostatic chemical systems[52] or chemo-mechanical oscillators[29], demonstrate intrinsic coupling where the sensed internal state directly drives actuating processes.

**(iv) Predictive capabilities through self-regulation** can emerge when the material's intrinsic physical dynamics and internal feedback mechanisms embody anticipatory processes, without resorting to external algorithmic prediction. Essential functionalities include:

***a) Anticipatory State Dynamics.*** The material's internal state evolution, governed by its physical laws and interaction history, can implicitly predict likely future configurations or environmental interactions. For instance, viscoelastic materials subjected to repeated stress cycles can develop an internal structure (memory encoded in polymer conformations or spring rest lengths) that reflects the history of applied loads. Subsequent relaxation dynamics are then guided by this acquired structure, causing the system to more rapidly find previously visited low-energy states, effectively anticipating stable configurations based on past experience[53]. Similarly, the internal memory state of the slime mold *Physarum polycephalum*, encoded in its tubular network hierarchy after encountering a nutrient source, guides its future migration towards that location, embodying a prediction of resource availability based on past encounters[54]. Active matter systems operating near critical points may also exhibit anticipatory dynamics, where long-range correlations allow the system to "sense" impending global changes based on local fluctuations[48,55].

***b) Physical Goal Embodiment and Error Correction.*** Self-regulation often involves maintaining a system's state around an intrinsic physical setpoint (e.g., minimum free energy, target morphology, homeostatic chemical concentration). Deviation from this setpoint acts as an implicit prediction error, driving physical dynamics that reduce the discrepancy. Self-healing materials exemplify this: sensing damage (a deviation from the intact state) triggers corrective processes (e.g., reactant release, chain diffusion) aimed at restoring the original low-energy, structurally sound state[29,56]. Theoretical frameworks like Active Inference formalize this, suggesting that biological systems minimize prediction errors relative to an internal generative model representing preferred (e.g., physiologically viable) states[57–59].

***c) Learned Predictive Response through Material Adaptation.*** Material systems possessing intrinsic plasticity can adapt their internal structure or parameters based on experience, effectively learning a predictive model of interactions. Nanowire networks function as reservoirs whose dynamics encode input history, allowing trained readouts to perform prediction tasks; embedding the learning within the network itself via intrinsic plasticity is a key goal[10,12]. Biological sensory adaptation, such as in bacterial chemotaxis where the methylation state of receptors adapts based on ligand history, allows the system to predict future concentrations relative to the adapted baseline and respond primarily to changes[60,61]. Chemical reaction networks can also be designed or evolved to learn predictive responses based on input patterns[62,63]. This learned predictive capacity arises directly from the adaptation of the material's physical or chemical state.

### 2. Memory and Ability to Learn

For soft matter to exhibit cognitive functions beyond reactive behaviours, it must possess a physical **memory** and be capable of **memory and learning**. This component allows the material to retain information about past stimuli and adapt future responses, accordingly, enhancing its adaptability over time. Important criteria within this category include:

**(i) Memory Encoding and Recall**: Intelligent soft matter can encode and retrieve information based on past interactions, such as deformation patterns or exposure to particular substances. Materials could employ phase-change elements or deformation-based memory systems to store and recall specific states. Hydrogels can retain and forget information based on thermal stimulation, showcasing how soft materials can encode and retrieve information based on past interactions[64]. A thermomechanical local-probe technique has achieved ultrahigh storage density in thin polymer films, allowing data storage, retrieval, and erasure[65]. This concept is linked to memory and entropy generation in non-equilibrated polymers, where the stochasticity and fluctuations in polymer chain conformations retain historical information that impacts their current behaviour[66].

**(ii) Self-Regulation and Environment Adaptation**: By recognizing trends in environmental inputs, the material can build predictive models that enable it to anticipate and prepare for future stimuli, adjusting pre-emptively to maintain stability and resilience[52].

**(iii) Self-Repair and Learning**: Self-repair mechanisms, a key feature of adaptive materials, enable the material to mend itself and develop resistance to recurring stressors, effectively ‘learning’ from repeated exposure[67]. This capability is essential for long-term functionality in changing or challenging environments.

**(iv) Adaptive Pattern Recognition**: Through the ability to identify recurring input patterns, intelligent soft matter can dynamically adjust its behaviour or structure to align with predictable environmental cycles, enhancing its utility in applications where environmental conditions fluctuate periodically[68].

### 3. Actuation

ISM must interact with its environment in a meaningful and self-directed way, making "actuation" an inherent property of the material itself. This involves moving beyond the mere replication of rigid robotic functions by passive compliant components, toward generating “embodied actuation” with self-sustained actions driven by the material's intrinsic properties and dynamics. This approach enables adaptive and self-regulating behaviours that fundamentally differ from existing implementations.

**(i) Soft Robotic Actuation**: Current soft robotic actuators, while offering unprecedented flexibility and adaptability in shape and movement, often require complex external control systems and high power inputs[2,46]. Future directions in ISM actuation should minimize dependence on such external controllers by designing materials with inherent actuation capabilities based on their structure, exploiting self-organization instead of hard-coded actuator arrangements for predefined mechanical operations[2,69]. The emphasis should not just be on designing better "actuators" that perform motion under stimuli in preprogrammed patterns. Instead, the focus should be on designs that possess inherent self-actuation properties through the physical and chemical components of a material structure. In such designs, actions should emerge from the intrinsic properties of the system itself, rather than relying on pre-designed modules. These modules often use external electrical hardware to power motion via external control or require separate components for every functional task, which limits system integration. To address this limitation one could explore architectures where the system relies on self-actuation through its structure for targeted, dynamic responses based on self-regulation using basic, well-established, and robust physical principles for action output[31,70].

**(ii) Programmable Morphological Transformations**: Beyond simple actuation mechanisms that are mostly limited to single mode mechanical deformations and responses for basic motions and linear actuator displacements or volume expansion-contraction and/or stiffness switch, truly intelligent soft matter demands a greater repertoire of dynamic morphological transformations[64]. Future research directions must therefore seek to engineer materials capable of more sophisticated and controlled shape changes, with a wider range of accessible kinematic behaviours that are tuneable according to desired complex functional actions by external interventions. This will likely be based on the new and better use of reversible material transitions, such as liquid crystals with electric or optical control or magnetically actuated polymeric systems, where the material can deform into multiple shapes in response to external stimuli, rather than preprogrammed, single-output responses for each input signal[2,24].

The next level of material design, beyond current capabilities, requires implementing fully programmable shape morphing metastructures, especially using buckling transitions, deploying hierarchical designed systems with complex topology and geometries or nonlinear mechanical behaviours of soft materials and structures. This transformative approach can build on recent advances in material science and soft robotics including new methods that couple self-organization with programmable response patterns[7,40,71].

#### 4. Decision making and communication

The **decision-making and communication** capabilities of intelligent soft matter enable complex, coordinated behaviours and distributed processing across the material's network. This component allows for collaborative actions within and across materials, fostering autonomous operation in sophisticated settings. Key aspects of this component include:
**(i) Distributed Decision-Making:** This criterion involves decentralizing control across the material, allowing individual nodes or regions within the material to make local decisions that collectively influence the overall behaviour [72,73]. This distributed system ensures flexibility and resilience, akin to biological networks[74–76].
**(ii) Self-Organizing Communication Pathways:** Adaptive communication channels form spontaneously within the material's structure, enabling efficient signal transmission[54]. These pathways can evolve with usage, optimizing for quicker responses and reducing internal communication delays, potentially mediated by physical fields or network restructuring[12,31].
**(iii) Integration in networks:** When multiple intelligent soft matter units are present, they can interact indirectly via mechanical, electrical, or chemical signalling[44,77]. This interaction enables coordinated group responses[31,78,79], allowing the material to perform tasks that require cooperation or shared objectives[80], simulating social behaviours in biological systems[73].

## Concepts Defining Material Intelligence

Intelligent Soft Matter draws inspiration from and aims to implement several foundational principles observed in complex adaptive systems, enabling sophisticated functionalities (Fig. 2).
**(i) Self-organization**, manifests as the spontaneous emergence of ordered structures and patterns from local interactions, mirroring phenomena seen across scales from lipid bilayer formation to the crystallization of colloids[81]. This principle allows ISM to create functional architectures without direct external templating, offering robustness and adaptability.

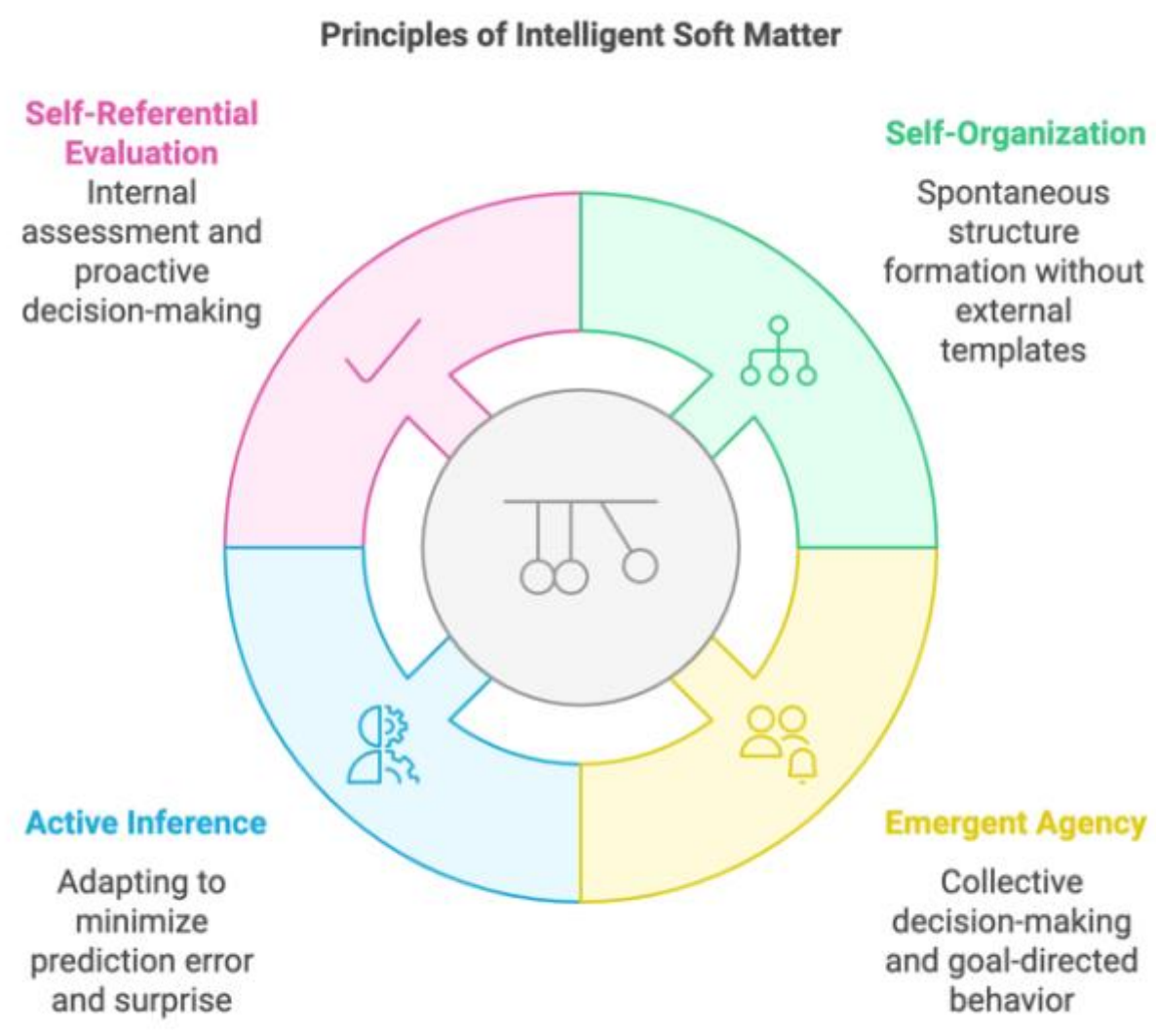

Figure 2. Four main principles of Intelligent Soft Matter

**(ii) Emergent agency** refers to how macroscopic behaviours, exhibiting goal-directedness or decision-making capacity, arise from the collective interactions of numerous simpler components. This is analogous to the coordinated movements of bird flocks[55] or the problem-solving capabilities of ant colonies[82], where individual units follow basic rules but the group accomplishes complex tasks. In ISM, this could manifest as a swarm of nanoparticles autonomously navigating a gradient or a self-healing material collectively repairing damage.
**(iii) Active Inference**[58] suggests a design principle for intelligent systems, including materials, whereby they shall operate so as to minimize surprise or prediction error. This is achieved by continuously refining an internal model of their environment and acting upon the world to validate this model. At a material level, this translates to dynamically adapting structural configurations or properties to better match and anticipate environmental changes. For example, a stimuli-responsive polymer network that swells or contracts in response to temperature can be viewed as actively inferring the environmental temperature and adjusting its state to minimize the deviation from its 'expected' state[83]. Thus, active inference provides a powerful theoretical lens for understanding how agentic systems (with a boundary between internal states and external sensory layer) can embody perception (sensing environmental cues), information processing (updating the internal model), and adaptive behaviour (modifying its state).
**(iv) Self-referential evaluation** posits a level of sophistication where materials not only react to external stimuli but also assess their own internal state and its relationship to the environment. This introduces a crucial element of intrinsic motivation and allows for actions that are not strictly stimulus-response driven. Drawing parallels with concepts in cognitive science like metacognition in biological systems or intrinsic reward mechanisms in reinforcement learning[84]. This suggests an ability for the material to evaluate its own performance or stability and make decisions based on this internal assessment. For instance, imagine a material that not only heals damage[56] but also learns from past damage events to proactively reinforce vulnerable areas, demonstrating an action informed by its 'awareness' of its own structural integrity and environmental stressors. While still largely theoretical in materials science, the implementation of self-referential evaluation would signify a significant step towards truly autonomous and context-aware intelligent soft matter, blurring the lines between engineered and biological intelligence.

## Challenges and Opportunities

The path to realizing the full potential of ISM is fraught with significant challenges, primarily revolving around material complexity, scalability, integration, and the realization of higher-order cognitive functions (Fig. 3). The most difficult challenges are related to overcoming current limitations of using specific designs with separated components for action (as in micro-bots), sensors (as seen in membrane-based chemical detectors), and memory (in the electrical or structural states of simple materials). Future implementations need to generate new architectures with true complexity to enable truly self-organizing systems with adaptivity. Any architecture must go beyond specific functional component units, such as single-modality-responsive sensors or hard-wired actuators. The primary goal is to develop materials that can exhibit higher-order capabilities. These include cognitive abilities such as dynamic responses, memory with feedback for learning, and adaptability at small spatial scales. Additionally, such materials should possess self-sustainability, moving far beyond pre-set responsivity. Such new directions call for exploring new types of design where information can be transduced and processed in an emergent manner, without relying on rigid hierarchical control. This can be achieved by designing all modular components to have a dynamic response based on coupled local processes specific to each element. These components should also interact in a more complex, highly integrated, and self-evolving manner. Instead of following a static and fixed response based on a predefined pathway, they should adapt and evolve beyond what is currently demonstrated.

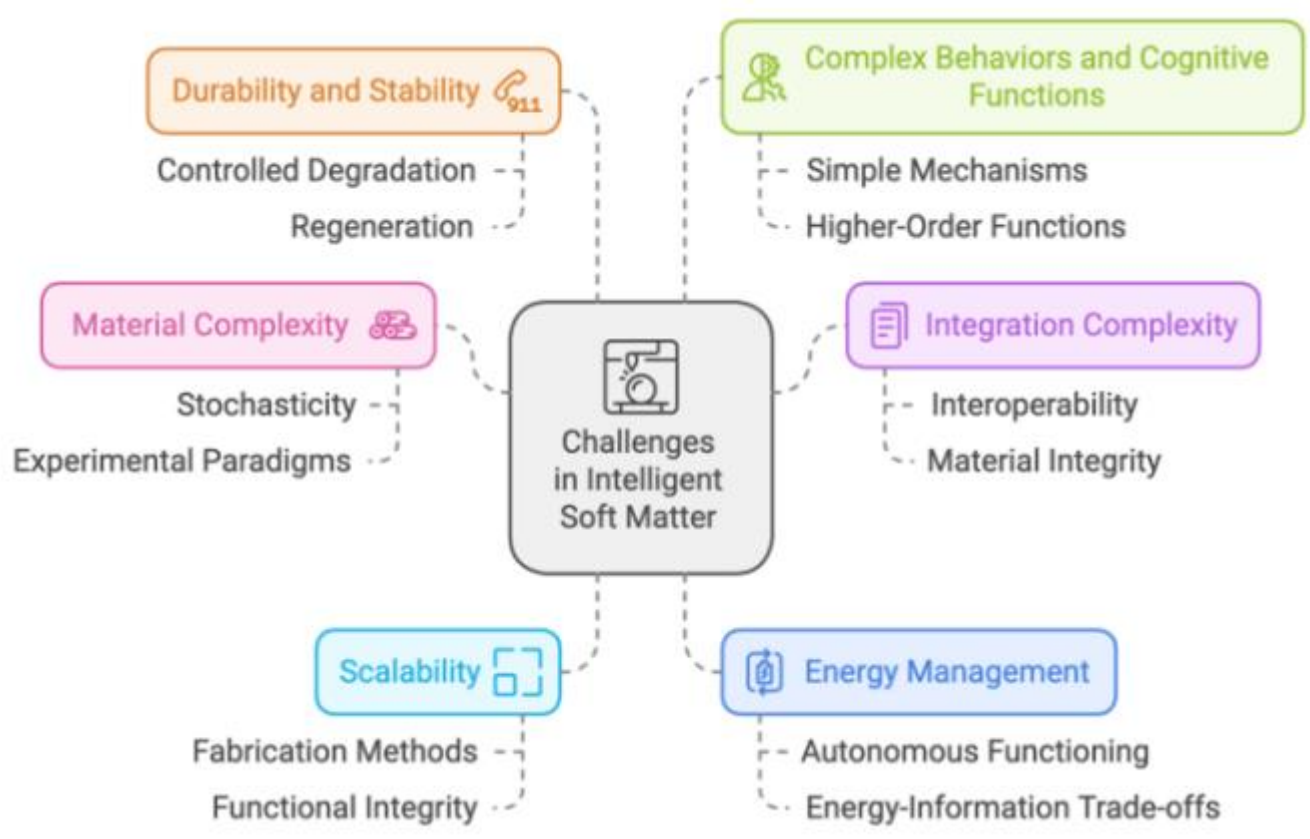

Figure 3. Challenges and opportunities in design of Intelligent Soft Matter

**Complex Behaviours and Higer-Order Cognitive Functions**

While current implementations demonstrate some form of functional behaviour, such as mechanically active robots[2], self-sensing membranes[32], and memristor-based circuits[9,85,86], these systems often exhibit behaviours which have predictable nature. Thus, they lack the features necessary to implement more than one set of functional operations via a rigid pre-set pathway and cannot possess abilities that are inherent to complex cognitive mechanisms. Present materials often rely on external electrical signals to switch response patterns rather than to control them by intrinsic properties of the material and its internal dynamic states. Therefore, most currently used experimental setups in ISM respond in pre-programmed manner to specific environmental changes or stimulus-action-based feedback using an implemented design that limits complex behaviours. Moreover, existing experimental systems often lack the integration of essential properties, such as self-evaluation and the selective processing of information, which are fundamental, for example, in living systems.

This raises a question regarding the definition of "material intelligence." Existing prototype devices, often categorized as "intelligent materials," are more accurately described as 'smart' or 'actively responsive' entities. While demonstrating sophisticated responses to specific stimuli and exhibiting complex functionality (dynamic shape change, memory storage, signal processing, and accurate actuation), these implementations are constrained by their dependence on external control and preprogrammed mechanisms, which prevent the demonstration of fully adaptable responses[2,9,31]. These limitations arise from a lack of intrinsic feedback, non-linear dynamics, and adaptive or self-regulatory properties. A critical shortcoming is the absence of mechanisms for internal selection or adaptation based on previously stored information. As a result, these systems exhibit predetermined pathways that lack dynamic interaction between the material and external inputs beyond what has been pre-programmed through external protocols, such as light signals or pressure patterns.

Thus, it will be paramount for future investigations, to focus on design frameworks which have at their basic operational level a built-in, local feedback control with the ability of modifying their parameters over time based not on external intervention but rather by integrating sensory information and actions as a single system with local functional parts. For example, when a stimulus is detected by a sensor, rather than relying on some pre-programmed rule to dictate the system responses, the sensor itself must respond with a self-adapting behaviour by actively changing its internal parameters or states, while maintaining structural integrity and robust responses over longer timescales. Such processes are employed by living organisms (for instance, when sensing and adapting with local memory, coupled with continuous energy harvesting).

Similarly, living biological systems perform at their best by an integration of external information with local internal adaptation[56]. For such purposes the building block components themselves need to possess many degrees of freedom, and integrate multiple functional behaviours within the smallest possible scale units by proper organization of its materials, composition and self-assembling processes[20,32].

Future frameworks must explore mechanisms beyond relying solely on classical electrical or optical signalling pathways, which often require centralized processing units and substantial external energy inputs. This approach is often inefficient and creates bottlenecks for truly autonomous operations that could occur also by coupling nontrivial forms of information transduction in micro components. Therefore, future research should prioritize designing systems with minimal dependence on external power or central controllers[87,88]. An emphasis must be placed on *locally driven processes* that leverage on inherently available resources from a variety of internal material or physical and chemical properties rather than complex external or added components.

This pathway encourages exploration of signal propagation based on material-based processes which directly transduce a stimulus into a set of actions or a memory with time dependent characteristics for autonomous functional devices. We also must exclude the need for externally powered/controlled digital/software components that rely on additional dedicated systems which typically consume a lot of energy for their operation and processing.

Another aspect is the criticality which can be an essential part of the design. Operating near criticality is often assumed to be advantageous for efficient computing[48,89]. However, simpler systems functioning outside the critical regime can be more easily controlled and may still offer sufficient computational capabilities for many applications[90,91]. Consequently, future research should investigate the role of criticality in intelligent soft materials and determine whether it is a fundamental requirement or simply one approach among many.

Transitions between ordered and disordered states of cognizant systems are of considerable importance also for explicit models of neural connectivity. There, one may ask how high sensitivity for input patterns and the ability to switch focus between tasks can be achieved[92]. There are indications that the optimally trained state is found between an overly ordered state (overtraining) and a noisy state (undertraining), referred to as the "edge of chaos"[93–95] by means of a Frobenius norm-based criterion. The boundary between dominance of order and noise is close to the optimally trained state[96], where the neural network's prediction is most informative for any new data presented at the inputs.

New architectures and fabrication techniques should create devices that display emergent behaviour while optimizing and minimizing the energy needs, material requirements and complexity. It should be an implementation of a real internal logic that directly impacts the material's physics rather than the translation of sensor output via algorithmic calculation.

The proposed method to achieve this high-level cognitive function demands exploration of new directions that involve designing a material by taking its self-organization dynamics with local energy sources and memory not as mere input-output transduction parameters or isolated functionalities of distinct units, but by making all these aspects as intrinsic components.

Our focus here extends and refines classical views on embodied intelligence[97,98] emphasizing that the physical body and its interactions are not mere peripherals but integral components of cognitive processing. It highlights the potential for intelligence to emerge even more deeply within the material substrate itself, potentially minimizing or even eliminating the need for a distinct, separable 'controller' or 'brain'-like component. While classical embodied intelligence often analyzes how a given (often biologically inspired) morphology facilitates computation performed elsewhere (e.g., in a neural controller), we are specifically interested in scenarios where the material's physics performs the computation, stores the relevant history (memory), and drives the adaptive response intrinsically[1,63].

The evolution (learning) in such structures should not require direct external intervention or external instructions but instead rely on local feedback loops. Thus, in these systems, intelligence must occur through the material's inherent nature and physics, and can be considered as "material-based intelligence" using embodied strategies and autonomous behaviours based on the system's physics as key components for complex responses.

#### Material Complexity

Biological systems self-assemble precisely and reproducibly into a target structure to fulfil their functions. Examples range from proteins to nucleic acids, membrane channels, and vesicles. This self-assembly process is a remarkable result of a wide range of reversible interactions (hydrophilic-hydrophobic, electrostatic, hydrogen-bonds, van der Waals, etc.), each with very different energy scales and ranges of action, all of which conspire to achieve a unique and remarkably precise result. The holy grail of material science nowadays is to attempt to reproduce this process artificially across scales. Figure 4 illustrates several examples of soft matter systems that conform certain aspects of ISM.

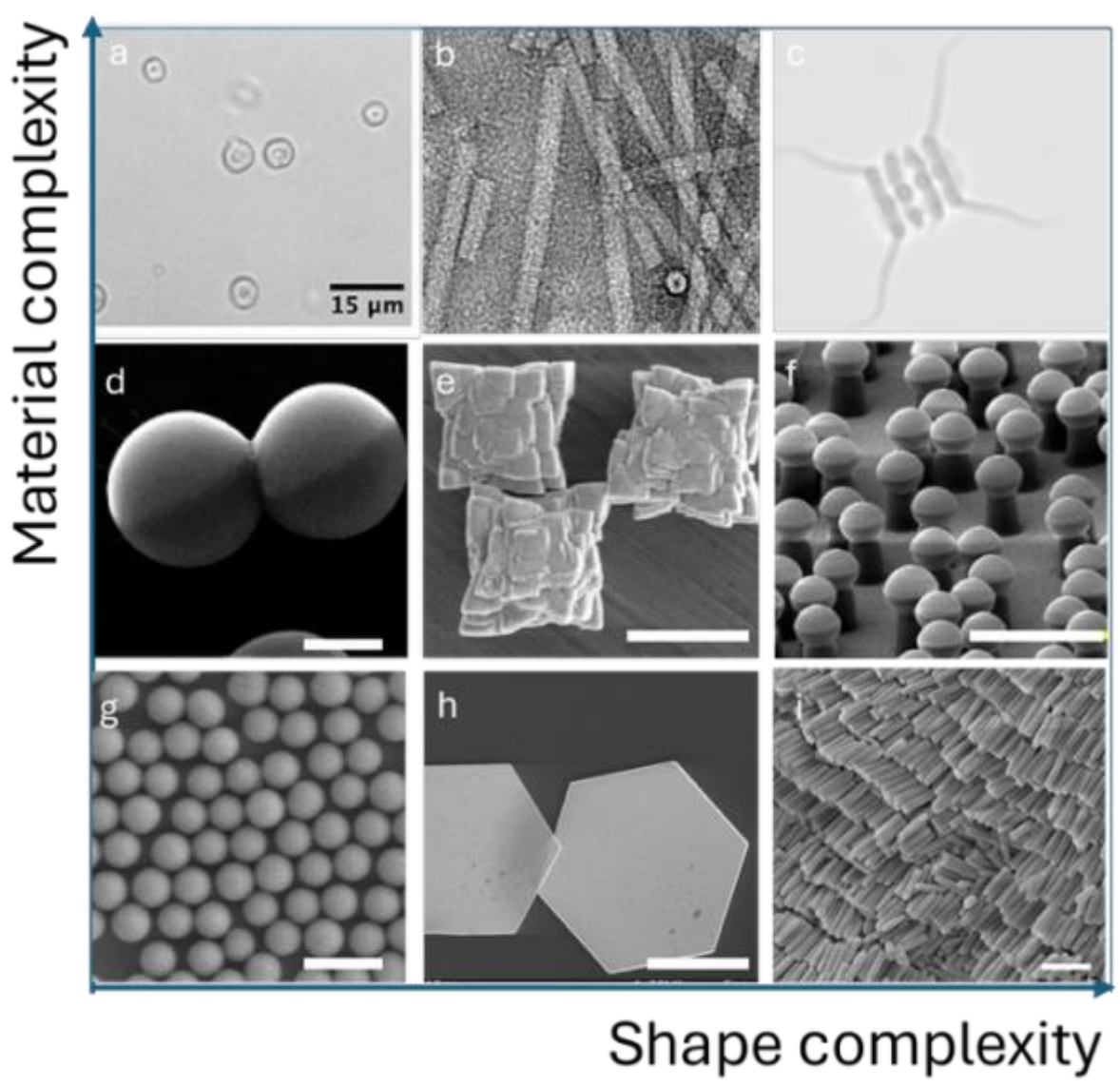

Figure 4. Matrix of visual impressions of material and shape complexity: (a) Polyelectrolyte filled non-biodegradable polyelectrolyte capsules in optical phase contrast microscopy see ref. [99]. (b) TEM image of tobacco mosaic virus (c) optical micrograph of a microalgae (d) SEM micrograph of Janus Au/ Pt colloidal particles (CNRS), scale bar 1 mm. (e) SEM micrograph of BiVO4 particle (f) SEM image

of Nafion based nanopillar arrays, see ref. [79] (with permission from the Royal Society of Chemistry). (g) SEM micrograph of SiO2 spheres (h) SEM micrograph of Au plates (i) SEM micrograph of assembles iron hydroxide rods, unless otherwise indicated, all scale bars are 2 mm.

Most existing systems exhibit limitations regarding the precise spatial organization of multiple, varied building elements that may work together in a synergistic way to obtain multi-functional behaviour, as it is seen in biological systems[2,2,70].

**Hierarchy**: Traditional engineering tends toward homogeneity within materials, and while hierarchical systems exist, they are usually built via top-down manufacturing by assembling components from diverse origins based on pre-determined functional units rather than exploiting synergistic properties emerging at smaller scales using bottom-up approaches. Often, they rely heavily on digital signal control, using specific types of electrical/optical/mechanical interfaces. While this enables high functionality and performance, it can compromise system efficiency or restrict device behaviour by a rigid structural design. Therefore, the challenge of achieving material complexity lies in designing systems from their core structure to enable higher-level self-assembly with minimal components, through simple chemical and/or physical couplings at the molecular level, leading to the emergence of system-wide coordinated dynamics rather than a complex set of steps. Such a process is based on the exploitation of existing material properties, rather than the introduction of external controllers and operations performed by separated units. It involves rethinking architecture to focus on integration rather than connection.

Yet, some indications of simple systems that may pave the way along these lines do exist and are well known. For instance, amphiphilic block copolymers[100] represents a simple paradigmatic example of a subunit that can be functionalized to achieve a target structure. Even simple synthetic polymers appear to share some similarities with biopolymers[101] and their self-assembly seems to obey some universal phase behaviour[102].

**Liquid Environments and Hydrodynamic Interactions.** In many soft-matter systems, particles are suspended in a liquid. This has important consequences for the dynamics, even in passive suspensions. However, in active, non-equilibrium systems it also affects structure and self-organization, in contrast to passive systems, where the structural properties are independent of the dynamics, a fundamental principle of statistical physics. In intelligent active systems, hydrodynamic flows and interactions affect the behavior in many ways[103–105]: (i) Propulsion flows can lead to repulsion or attraction of neighboring microswimmers (an example is the hydrodynamic starvation of fish larvae, as hydrodynamics pushes away the food they are trying to capture). (ii) Steering flows, which are required for active reorientation, can modify the motion of neighboring particles. (iii) Hydrodynamically generated torques can change the orientation of nearby particles (implying, for example, that particles cannot move cohesively through alignment steering alone, but in addition speed adaptation is also required). (iv) Generation of strong fluid jets, swirls, and active turbulence due to clustering and swarm formation of many self-steering particles. This poses significant challenges for intelligent agents navigating and maneuvering in liquid environments, as they must overcome adverse hydrodynamic effects and instead aim to exploit hydrodynamic forces to their own advantage.

**Functional integration:** A key challenge in achieving true material complexity is integrating essential capabilities (Fig. 1) within a single unit or a small group of component units. These units must contain multi-stimuli-responsive parts while enabling synergistic effects between them. Instead of relying on sequential, non-coupled blocks, the system should facilitate different types of energy transductions. This approach ensures that the output or behaviour of one part directly influences another at its core without requiring additional systems such as controllers, feedback mechanisms, or computers. Such functional integration is rather complex and may require creation of novel chemical or physical interfaces where materials mutually influence each other for more complex information exchange[9].

Current experimental realizations of complex behaviour are typically achieved through the assembly of independent units[72,106–110], each contributing a well-defined part or step in the overall cycle. This creates devices that perform operations or behaviours in a step wise manner at discrete spatial and temporal points but lack the ability to dynamically couple behaviours through self-organized processes, as is commonly observed in living matter, which features inherent feedbacks and cross-communication capabilities[56,88].

**Beyond centralised control**: Future designs should also explore mechanisms by which different responses are triggered by external parameters with graded behaviours rather than binary states. This could involve local signal processing units where the output of one stage will become the input of the other using a network structure with multiple interactions and pathways to modulate responses. In doing so, it would emulate important aspects of biological material behaviour, which rely on coupled information pathways to achieve higher performance while minimizing noise from the local environment[111].

We need new architectures that explore designs where all structural components contribute to functionality, rather than relying on specialized functional units. This approach would reduce the number of components needed to achieve functional complexity, allowing shared functions to emerge from the architecture itself. Additionally, local nonlinear responses to various stimuli should be achieved by altering the material's internal structure, which in turn changes the system's behaviour. Finally, the design should move beyond singular response mechanisms to develop multimodal responsive materials, not only by acting and changing mechanical or electrical properties based on a specific trigger but also responding through the all-encompassing variety of material capabilities when exposed to multiple simultaneous types of triggers. Therefore, an important criterion would be creation of material systems that can respond through multiple modalities such as light, temperature, mechanical or chemical stimuli, all arising from similar or interacting structures that act on local parts of the material itself[32].

A future direction must also prioritize designs, which promote new levels of dynamic changes with both spatial or temporal features to create versatile functional behaviour rather than discrete changes with limited scope. Furthermore, these functional parameters should appear spontaneously without explicit external programming or guidance, *i.e.* the system should show the capacity to create dynamic configurations that can also self-regulate locally, maintaining robust, efficient and predictable operation while performing sophisticated behaviours[20,31].

#### Energy Management

One of the definitions of embodied intelligence as a goal-directed exchange of energy and information[112] suggests that it is fundamentally grounded in the physical processes of energy transduction, utilization, and information acquisition required to achieve goals within environmental constraints. As explored in recent works[113,114], the thermodynamic costs associated with sensing, computation, actuation, and adaptation are not merely overheads but fundamental constraints shaping intelligent behavior. Biological systems, driven by evolutionary pressures, exemplify extreme energy efficiency, providing a benchmark and inspiration for ISM. Therefore, designing ISM requires moving beyond simply supplying power towards optimizing the entire energy workflow from harvesting ambient energy to minimizing dissipation during information processing and maximizing useful work output, all within the framework of achieving the system's objectives.

**Energy harvesting**: Energy autonomy stands as a critical yet often under-addressed factor in current ISM prototypes[36,115]. While there are examples of energy harvesting materials at small scales such as those converting light or chemical energy into driving forces for micro robotics, it is generally inferred that energy efficiency is an advantage of minimal computational units[70]. However, most research on ISMs has not properly quantified the energy requirements for system operation, nor have these systems been explicitly designed to harvest power for perpetual operation. This limitation is notable because, in most instances, present ISM implementations are heavily dependent on external energy sources to power their actuation, sensing, and computational processes[9,88]. This external reliance not only limits scalability for real-world deployment but also inherently undermines true autonomy, particularly in situations requiring long-duration operation in remote locations or resource-constrained scenarios. A fundamental shift in design approach should focus, instead, on integrating self-powering mechanisms directly within the material architecture. By analogy, in living organisms, self-sufficiency for energy needs and homeostasis is a fundamental requirement for their survival in absence of continuous and direct "human control"[56,69]. Therefore, any practical implementation of "cognitive and adaptive characteristics at material level" requires incorporation of efficient processes that harness energy directly from the environment without externally supplied energy sources that pose major operational limitations[9,88]. For example, biological systems can harvest energy from their environment through photosynthetic or oxidative mechanisms.

**Non-equilibrium states**: A critical aspiration for advanced material intelligence, is the ability to maintain complex organization and function persistently far from thermodynamic equilibrium[116,117]. This requires a continuous energy throughput to counteract the universal tendency towards decay and disorder dictated by the second law of thermodynamics[35,118]. Consequently, any material system exhibiting autonomous adaptation, computation, or self-maintenance must operate as an open system, harnessing energy fluxes to sustain a non-equilibrium steady state (NESS) or to engage in dynamic, non-equilibrium processes[119].

However, maintaining NESS and performing functional tasks incurs thermodynamic costs, fundamentally linked to energy dissipation and entropy production[34,36,120]. This presents a significant challenge and a key differentiator between current "smart" materials and the vision of cognizant matter. Many contemporary active or responsive materials, while functional, are often highly dissipative or energy-intensive. For example, memristive and nanowire networks consume energy during state switching and readout[10,121,122]; electrothermal actuation in soft robots often suffers from significant heat loss[51,123]; and chemically propelled microswimmers typically exhibit very low chemo-mechanical energy conversion efficiencies due to viscous dissipation at low Reynolds numbers[70,124,125].

Furthermore, many current systems operate robustly only within narrow, externally maintained parameter regimes. The trade-off between the energy required for processing (e.g., switching a state, performing a computation, achieving a certain accuracy) and the robustness or reliability of the outcome often limits the adaptability and broad applicability of these systems[87,126]. This reliance on precisely controlled external conditions contrasts sharply with biological systems that maintain function across wider ranges, often by dynamically regulating their internal state and energy usage. Therefore, a crucial direction for advancing material intelligence is the development of materials and systems capable of achieving optimally working NESS. This implies minimizing unnecessary energy dissipation while maximizing functional performance (*e.g.*, computational accuracy, adaptive capacity, robustness) within the constraints imposed by physics.

**Active local units**: Future design should also consider implementation of active, local units that serve as energy transducers of locally available chemical and mechanical energy and rely on internal dynamic mechanisms of materials for core operations instead of using any active external control components with their specific hardware requirements (such as batteries and microcontrollers)[61,127,128]. Ideal targets for the design of intelligent soft matter systems should focus on minimal energy designs. These materials should be capable of exploiting gradients or other environmental energy sources, such as mechanical stress, light, temperature variation, and chemical/pH gradients. This energy would be used to power memory storage, computation, and actions.

For instance, in active photonic-based materials, coupling with light energy can enable multispectral light harvesting and wave modulation. Newly designed materials with self-organized structures may convert radiation from diverse wavelengths into localized internal energy, while also exhibiting dynamic material-based behaviours[129]. This approach eliminates the need for external electrical sources to deliver energy to localized regions or to trigger specific actions. Instead, energy supply becomes an autonomous integrated process driven by internal chemical, thermodynamic, or photonic mechanisms embedded as internal building blocks within the material's architecture.

As a result, we must go beyond existing concepts of actively driven materials to those that extract power from their environments. Power extraction must also occur locally, with appropriate management of energy production (chemical, optical or electrical), transport, and consumption at the micro level of an intelligent material, rather than relying on external devices coupled to an electronic system for power.

**The interplay between entropy and information processing** is central to understanding the emergence of intelligence in soft matter. Entropy production, traditionally a measure of disorder, is reformulated as a driving force for self-organization, in direct relation with information, where the processing and storage of information become thermodynamically relevant operations[87,130]. As highlighted by fluctuation theorems and minimal models in stochastic thermodynamics[58], minimizing entropy production – or, dually, maximizing entropy under constraints–becomes a crucial principle in understanding the energetic costs associated with accurate and reliable information handling. In polymer systems, for instance, non-equilibrium states, representing a form of a "memory", can be understood as states of reduced conformational entropy, stabilized over long timescales, reflecting the system's past history and interaction with external parameters such as shear rate, evaporation rate, mechanical stress or temperature[8,66,131,132]. Exploiting the inherent link between information and entropy could lead to the development of material-based computational platforms. These platforms would enable efficient, low-power data processing in complex and dynamic environments. Instead of relying solely on algorithms, they would harness their own physical and chemical properties for adaptation and optimized responses. These responses would be intrinsic to the system's architecture rather than merely externally added or pre-programmed elements.

Sensing, recording, actuation, and communication (Fig. 1) in any intelligent system are inherently thermodynamic processes that necessitate a continuous flow of information and energy[126]. These fluxes must be carefully balanced, or "budgeted" at the mesoscale levels within any given design, for the functional material system to efficiently transduce, propagate and utilize energy and information within each operation cycle and to avoid undesirable dissipation for scalable low-power implementations. Such balance is naturally achieved in living systems through metabolic pathways[69], where energy fluxes are optimized to sustain cellular functions. Therefore, any new generation of "cognizant" material systems will also need to integrate these fundamental rules if truly high efficiency, scalable and robust technological outcomes are expected---mirroring, in that context, a more natural biological blueprint.

For example, quality-dissipation trade-offs[128] represent a fundamental principle for designing energy-efficient and reliable ISM systems[36]. According to thermodynamic uncertainty relations, improved functional performance, such as enhanced sensing accuracy or faster information transfer rates, must be compensated by increased entropy production (dissipation) and higher energy expenditure within the system[88]. This implies an intrinsic thermodynamic cost associated with "intelligence" of material operations, particularly in non-equilibrium conditions. It highlights the critical importance of energy optimization strategies in ISM design, where the balance between computational accuracy, response speed, and energetic cost determines the efficiency and practicality of such technologies[36].

#### Integration Complexity

Integration complexity, along with material compatibility is a considerable limitation to further implementation of complex and truly self-organizing materials. Most devices lack hierarchical component organizations and a clear control of component interface (*i.e.* a high level of architectural heterogeneity using multiple functional parts) by usually depending on uniform structures with limited types of signal transfer and without spatial organization. To enhance such integrated behaviour, we need materials design strategies where structural elements, as well as their functions and properties, are combined through new design rules that do not require rigid pre-determined assembly protocols, allowing for an easy production or design of complex architectures. Further limitations stem from separated approaches where sensors and actuators have specific designs and are put together in a single device, leading to bottlenecks in performance due to their material/electrical compatibilities or due to lack of integration of a single "active element" (*e.g.* material system which possesses all properties, not separately created in multiple isolated areas)[2,58,133]. Instead, to have better outcome in terms of integration at a system level, such architecture requires creating modules with all functions in same building blocks. Ideally each part of an integrated unit would have also their own inherent mechanisms to change its structure and adapt for different functions locally. For example, this could be a set of molecules (proteins, peptides) creating tuneable interactions among themselves that act like integrated circuit-based logic but all are coupled by material design and responses, that should also act as transducers with minimal external parts or components to be wired or controlled using external computation resources. The current implementation strategies that rely on multilayer structures or multi steps are also challenging in their manufacturing throughput and system fragility or limitations due to component complexity for large-scale implementation. Furthermore, such limited degree of component interactions reduces greatly the potential functionality of such system if compared with biological models where different protein or DNA chains with multiple levels and degrees of connections are employed. We need integration without compromising any individual functional performance in any module but with their mutual connections playing a new role by a modular approach that gives versatility to the whole design paradigm.

Several examples of integration and material compatibility have been realized. For example, the colloidal assembly strategy allows for the integration of different functional agents and responsive mechanisms within a single system, resulting in multi-functional colloidal assemblies for complex tasks such as drug delivery[134–137] or elasto-active structures[138] leveraging nonlinear elasticity for directed movement.

Hierarchical structures, such as fractals, have been demonstrated to enable multi-band operations in antenna design[139]. This principle extends to plasmonic sensors, where fractal designs of plasmon nanoparticles have been extensively explored to achieve multi-wavelength sensing capabilities[129,140,141]. Analogously, fractal nano resonators exhibit[142] a broadband frequency response spectrum with high quality factor resonance peaks. These localized vibration modes can potentially function as multi-band mechanical sensors, capable of simultaneously measuring force, mass, or chemical compounds. Furthermore, the broadband spectrum of these mechanical fractal resonators opens possibilities for multi-band energy harvesting. Fabricating such devices using CMOS piezoelectric materials, as is achievable with Piezoelectric Micromachined Ultrasonic Transducers (PMUTs)[143], also allows miniaturization of these devices. The feasibility of this concept has also recently been verified at the macroscale[144,145].

Integrated sensing and actuation allow materials to respond dynamically to environmental changes, enhancing their adaptability and functionality. Precise manipulation of soft matter units, such as cells and microdroplets, can be achieved with active dielectrophoretic (DEP) forces. Spatial and temporal modulation of the applied electric fields allows real-time control over particle movement, enabling tasks such as sorting, trapping, and assembly within microfluidic devices[146]. Insulator-based dielectrophoresis (iDEP) facilitates particle manipulation without the need for embedded electrodes, making it particularly valuable for bioengineering applications, including diagnostics and tissue engineering. A critical parameter for fine mechanical control in these systems is the zeta potential, which depends on the surface charge of particles and influences their stability, aggregation, and interactions between different phases in the system. Accurate measurement of zeta potential is essential for designing and optimizing intelligent soft matter building blocks[147]. Electrokinetic actuation could significantly expand the potential for creating intelligent soft matter capable of reacting and adapting to dynamic, noisy environments.

Another platform for programming decision-making in autonomous soft matter is the liquid droplet system that can be designed to be sensitive to various environmental signals such as pH and salt gradients and translate this external information into droplet motion[148,149]. Both passive droplets that purely respond to external stimuli through chemotaxis[150] and active droplets that contain chemical potential for autonomous motion are good examples of the integration of sensing and actuation. The next steps in developing these system would be to demonstrate decision-making in the presence of various stimuli and changes of internal state of the droplets[148,151].

As mentioned, chemically triggered mechanisms are key ingredients for conferring autonomy to micro- and nanosystems. For instance, catalytic reactions occurring at the surfaces of nano and micron scale soft matter particles suspended in fluid, provide various mechanisms for the production of rapid motion, which rely on self-generated gradients[70]. It is well established that the details of the catalytic reaction, the catalyst distribution, the overall shape and size of the soft matter, and the surrounding environment properties can be used to control the type of motion produced. Demonstrations include the ability to engineer the relative amount of translational and rotational thrust, bias motion with respect to gravitational fields, and exploit topographical guidance[152]. External stimuli such as magnetic fields can alternatively be used to steer catalytically propelled devices in 3D paths, but with reliance on external actuation and control[45].

Additionally, new scenarios for chemical gradient activation are emerging in the field, such as polymeric micro/nanosystems activated by ion-exchange reactions[127]. These approaches offer alternative strategies to catalytic methods, which can sometimes be limited by issues like salt tolerance or the availability of innocuous fuels, especially challenging for bio applications[153]. For instance, it has been demonstrated that asymmetric ionomeric micro/nanostructures are active, regenerable swimmers that are activated by salts. These structures can interact with one another, generate gradients, sense local conditions, self-organize, adapt their velocity, and form collective behaviors, such as motile swarms that change speed in response to their size and shape[79]. These systems have the advantage of being versatile in terms of manufacture, size, and shape, and they can be modularly coupled with different inorganic, organic, and biomaterials to enhance their sensing capabilities and multifunctionality. Furthermore, they can be equipped with components that induce oscillatory reactions, which provide dynamic regulatory and feedback mechanisms. Oscillatory reactions can generate spatial or temporal patterns, such as chemical waves, which can be harnessed for encoding information or guiding the nanomotors in specific tasks. For example, periodic changes in chemical gradients could control propulsion speed, directionality, or interaction strength with other particles. By coupling with shared oscillatory fields, multiple nanomotors could enhance collective behaviours, such as synchronized oscillations or dynamic clustering. This coordination could lead to emergent intelligence, enabling the system to exhibit complex problem-solving or adaptive behaviours.

Emergent dynamics and self-organization in ensembles of self-steering cognitive active particles[154,155] addresses the question how the properties and interactions of individual cognitive particles – such as vision-guided pursuit, parallel alignment with neighbours, and resulting steering torques[78,156,157] determine the cohesion and collective behaviour of crowded systems and swarms[81,156,158]. It will take some time to design and engineer microbots which have all these functionalities, and "millibots" are more likely candidates where this can be achieved in the foreseeable future, but it is important to explore types of emergent collective behaviours in simulations now, in order to provide guidelines for microbot design. Ensembles of active self-steering particles can self-organize into dynamic structures that exhibit cognitive functions, such as learning from task performance and adaptation to changing environmental conditions. The emergence of such coordinated functionalities from simple constituent behaviours underscores the potential of decentralized intelligence (also called distributed computing) in achieving sophisticated material responses without centralized control.

In the context of intelligent systems, navigation would ideally be performed autonomously, without external guidance, to enable soft matter devices to identify and seek their own target locations[128]. While the full manifestation of this goal may require incorporation of memory and processing capacity, it is interesting to explore the limits that can be achieved using only a responsive and local environment sensing material. One example is catalytic micro-swimmers, that expand and contract in response to local pH variations. These devices were shown to be able to autonomously accumulate in low pH regions and autonomously modulate the release rate of encapsulated cargo[159]. This mechanism can enable devices to follow a stimulus that moves and varies in strength with time, and so provides a useful, responsive navigation capacity. Envisaged augmentations to this system include incorporating autonomous temporal responses via enzymatic "clock" reactions and using variations in the encapsulated cargo release rate to facilitate collective behaviour via intra-device signalling, in analogy to quorum sensing.

**Scalability**

A crucial aspect for the next step in material intelligence is scalability and how to keep device operational ranges in extended scale with a reliable architecture design. Practical challenges in scaling include limitations of top-down nanofabrication methods that often demand expensive cleanroom resources, hindering high-throughput production of systems at nanoscales[40]. Mass production and consistent replication of complex, hierarchical architectures with a large number of interconnected components and precise control over local properties and geometries remain a major obstacle to overcome.

Self-organization-based techniques that rely on basic physical or chemical driving forces could address this challenge by allowing building blocks to spontaneously arrange through interactions at the component level or with the environment, leading to macroscopic behaviours without many intermediate steps during their fabrication[24]. While self-organization techniques offer pathways to scalable manufacturing, achieving reliable and reproducible control over emergent behaviours and material qualities at large scales is another set of challenges for scaling that must be solved[98].

Scaling down to micro- and nanoscales introduces further fundamental and technical constraints, related to the increasing influence of thermal fluctuations and stochasticity at smaller dimensions, which are often neglected or simplified at larger scales. At reduced scales, standard methods based on classical, digital electronics for precise control, and high energy efficiency may also become less viable. In these regimes, quantum effects, molecular-level interactions, and surface forces become dominant, demanding new fabrication methods that must go beyond current high-resolution printing and nanofabrication limits[9,27].

To build scalable material designs, new minimal units that can interact by using common mechanisms rather than very limited single-component-specific interactions should be favoured, with an emphasis on modularity with a simple set-up procedure without the need of calibration with specific environmental control (*i.e.* use more natural physics/chemistry phenomena as much as possible as control mechanisms). A good implementation should aim at low resource requirement such as: using inexpensive and robust (stable at different experimental/operating conditions) starting materials and reactions, low temperature implementation, and a minimization of waste products.

**Durability and Stability**

Another important limitation for present soft "intelligent materials" is related to longevity or robustness during complex, diverse and variable operating settings. Often materials are fragile, have limited stability during environmental changes, or show fast degradation under physical or chemical harsh treatments.

Moreover, any realistic system should also possess a level of noise resilience for variations in its chemical and electrical properties due to long operations times or large volume systems. The new material implementation must then adopt approaches that create some form of implicit auto calibration of the materials by incorporating components with adaptive capabilities that account for system changes, such as using active polymer chain configurations (*i.e.* materials that adopt energetically favorable shapes and return to them even when perturbed ) for memory storage or in self-regenerating or healing matrices (*e.g.* self-assembling of material where the materials can self-organize and also produce components of their own matrix to restore previous structures). Locally available free energy should be used to maintain functionality under dynamic and variable conditions of operations to maximize system performance and stability during the full operation range without failure. In contrast, to achieve such robust behaviour, it is also necessary that materials or building units that will act as the "components" (sensors, processing or memory units) of this system possess inherent material-based properties that allow them to self-calibrate locally based on internal dynamical parameters without reliance on an external process.

For intelligent soft matter to be practically viable, future systems must extend beyond mere responsiveness and exhibit robust, self-improving, and stable operation over extended durations[56,133]. This necessitates a shift towards material architectures capable of self-regulation, wherein the system autonomously maintains functionality and stability in the face of perturbations, damage, and dynamically changing operational environments[20,35]. Such robustness might be achieved through exploring designs that draw inspiration from self-regulating biological systems[70]. For instance, materials with inherent structural self-regulation could be implemented via dynamic crystallization, self-limiting chemical reactions, or leveraging non-linear feedback mechanisms in phase-separated fluids and solid-state systems to ensure longevity and reliability without constant external adjustments[24,160]. A central challenge lies in developing materials with inherent physical properties that robustly respond to external and internal signals in a consistent manner over long time periods while maintaining their functional integrity without degradation, much like biological systems maintain homeostasis even when facing complex and changing environmental pressures.

## Conclusions

ISM is set to transform materials science, not only by integrating cognitive capabilities traditionally associated only with living systems, but also by introducing groundbreaking potential for autonomous, adaptive, and self-aware materials. While current research predominantly focuses on macroscopic demonstrations of sensing, actuation, and memory, we invite the exploration of the unique opportunities that emerge at the nanoscale. We argue for a paradigm shift that embraces intrinsic stochasticity and fluctuations as integral design elements, propelling the development of intelligent soft matter capable of advanced functionalities by exploiting the huge numbers of interacting units available at the nanoscale. Unlike traditional materials, ISM moves beyond passive responsiveness towards dynamic, evolving functionalities, either by mimicking the sophisticated behaviour of biological systems, or by embedding synthetic intelligence within the material itself. The interplay of distributed processing, complex network topologies, and the inherent material dynamics of intelligent soft matter establishes the foundation for novel computational paradigms and versatile sensing and actuation functionalities, with no analogues in traditional approaches. This form of distributed control, combined with non-linear system dynamics, unlocks the potential for self-learning and adaptation, enabling a variety of cognitive behaviours to emerge without the need for a central control or "brain".

This perspective delves into the thermodynamic foundations, bio-inspired designs, advanced material developments, and computational intelligence that underpins this new frontier. By addressing current challenges and highlighting future directions, we envision a future where intelligent soft matter integrates seamlessly with our world, enabling transformative applications across diverse fields.

## Author contributions

All the authors participated in preparation, creation of the manuscript, specifically writing the initial draft.

## Conflicts of interest

There are no conflicts to declare.

## Data availability

No primary research results, software or code have been included, and no new data were generated or analysed as part of this perspective paper.

## Acknowledgements

The work is the result of the SoftComp Topical workshop on Intelligent Soft Matter, Salou 2025 (https://softmat.net/intelligent-soft-matter/) financed by SoftComp Network of Excellence (https://eu-softcomp.net/). Various AI tools were used for preparation of the manuscript: language models Google Gemini 2.0 series and Discovery Engine (https://explore-the-unknown.vercel.app) for literature processing, structuring contributions, finding concept overlaps and summarizing according to procedure explained in https://github.com/vbaulin/IntelliDE/.

## Notes and references